# Deterministic coupling of site-controlled quantum emitters in monolayer semiconductors to plasmonic nanocavities

*Yue Luo[1,2], Gabriella D. Shepard[1,2], Jenny V. Ardelean[3], James C. Hone[3] and Stefan Strauf [1,2\*]*

[1] Department of Physics, Stevens Institute of Technology, Hoboken, NJ 07030, USA

[2] Center for Quantum Science and Engineering, Stevens Institute of Technology, Hoboken, NJ 07030, USA

[3] Department of Mechanical Engineering, Columbia University, New York, NY 10027, USA

*Address correspondence to: strauf@stevens.edu

**Solid-state single-quantum emitters are a crucial resource for on-chip photonic quantum technologies and require efficient cavity-emitter coupling to realize quantum networks beyond the single-node level[1,2]. Previous approaches to enhance light-matter interactions rely on forming nanocavities around randomly located quantum dots[3–8] or color centers[9,10] but lack spatial control of the quantum emitter itself that is required for scaling. Here we demonstrate a deterministic approach to achieve Purcell-enhancement at lithographically defined locations using the sharp corner of a metal nanocube for both electric field enhancement and to deform a two-dimensional material. For a 3×4 array of strain-induced exciton quantum emitters formed into monolayer $WSe_2$ we show spontaneous emission rate enhancement with Purcell-factors ($F_P$) up to $F_P=1050$ (average $F_P=272$), single-photon purification, and cavity-enhanced quantum yields increasing from initially 1% to 15%. The utility of our nanoplasmonic platform is applicable to other 2D material, including boron nitride, opening new inroads in quantum photonics.**

Multi-node quantum optical circuits that combine quantum light generation, manipulation, and detection directly on a chip are intensely sought after with the goal to realize advanced computation and communication platforms. To this end deterministic assembly techniques are required that not only place quantum emitters into desired locations but also reliably couple their spontaneous emission (SE) to optical waveguides for photon routing as well as to cavity modes to utilize light-mater interactions in the weak and strong coupling regime. A number of approaches for deterministic cavity coupling involving randomly distributed quantum dots (QDs) were shown, including atomic-force microscope locating of tracer QDs with respect to markers[3], cryogenic scanning of exciton emission location with respect to markers[5,7], direct cryogenic lithography after locating randomly distributed QDs[6], and in-situ electron-beam lithography[8]. Deterministic coupling to photonic crystal modes or planar microcavities was also shown for nitrogen[9] and silicon[10] vacancy centers in diamond, that both form at random locations within the diamond host crystal.

Common to these approaches is a lack of spatial control of the quantum emitter itself that form naturally only in random locations, thereby limiting the implementation of more complex quantum networks. A very recent work demonstrates deterministic waveguide coupling to three randomly located QDs on the same chip[8]. Some progress has also been made with site-controlled growth of QDs with Purcell factors ($F_P$) up to 4 when coupled to dielectric cavities. [11,12] Recent advances in monolayer semiconductors based on transition metal dichalcogenides (TMDC) have demonstrated that tungsten diselenide ($WSe_2$) can host randomly located quantum emitters[13–16]. Towards their spatial control it was also shown that strain-induced excitons can be formed intentionally via nanobubbles[17] as well as by lithographically defined arrays of pillars that act as local stressors[18,19]. This unique ability to create strain-induced 0D excitons anywhere within a 2D material is promising towards scalable quantum technologies, but so far lacks mature cavity integration.

With the goal to enhance the light-matter interaction of strain-induced quantum emitters in TMDCs, plasmonic nanocavities are particularly attractive since they feature nanometer sized gaps with ultra-small mode volume resulting in drastically enhanced SE rates of adjacent quantum emitters[20–22]. Their small footprint of about 100 nm allows for denser integration as compared to their dielectric counterparts such as photonic crystals. In general, proximity to metals can result in dominant nonradiative (NR) exciton recombination, i.e. photoluminescence (PL) quenching. In contrast, nanoplasmonic gap-mode resonators feature a dominant radiative recombination when emitters are placed in close proximity[23,24]. Using vertically oriented plasmonic gap modes, large Purcell factors up to $F_P$=1000 were demonstrated[20,23]. The ability to provide large $F_P$ values opens furthermore the opportunity to enhance the quantum yield (QY) of quantum emitters that suffer from intrinsic NR, as we recently demonstrated for the case of single photon emission from carbon nanotubes by enhancing the exciton QY from initially 2% up to 62% [25]. Despite these appealing properties of plasmonic gap-mode resonators, they have so far only been applied to enhance the ensemble exciton emission from TMDC material[26], i.e. to classical light emission. One significant difficulty to enhance the quantum light emission is the typical approach to disperse colloidal plasmonic nanocubes randomly onto a substrate containing emitters, thereby lacking spatial control of the plasmonic hot spot with respect to the quantum emitter location[20,26].

Here we demonstrate a deterministic approach to create the desired exciton quantum emitter directly at the hot spots of a plasmonic gap-mode cavity. This is accomplished by using the sharp corner of a lithographically defined metal nanocube both for electric field enhancement and to deform a two-dimensional material, which enables local strain-induced quantum emission, as shown schematically in **Fig. 1a**. The nanocube edges form vertical plasmonic gap-modes against the planar Au mirror, theoretically giving rise to a 900-fold intensity enhancement (**Fig. 1b, Supplementary Note 1**). The gap size itself is solely defined by layer deposition rather than lithography. The $WSe_2$ monolayer is surrounded by 2 nm thick $Al_2O_3$ spacer layers on each side, resulting in a total gap size of 5 nm (**Fig.1a** inset). This spacer prevents optical quenching and spectral diffusion in the exciton emission[25]. **Fig.1c** illustrates that hot spots are located at the four nanocube corners. Their spatial locations coincide with the occurrence of high-strain areas in the stamped monolayer (**Fig.1d, Supplementary Note 2**). An important parameter to optimize the success rate of spectrally isolated quantum emitter formation is the height of the Au nanocube[18,19]. Best results have been found at near-unity aspect ratio of flat-top cubes with 90 nm height and 110 nm side-length. **Figure 1e** shows a scanning electron microscope (SEM) image of an Au nanocube array with 2.5 µm pitch (see methods). The small size of the nanocubes allows efficient extraction of the nanogap emission into the far-field[20], while the transparent sapphire wafer allows for probing quantum emitters even when the cavity is closed. The mode resonance was tailored to coincide with the exciton emission wavelength (750-800 nm). As depicted in **Fig.1f**, the resulting broadband response (Q~8) allows the spectral resonance condition between emitter and mode to be automatically fulfilled without the need for spectral tuning, as is typically required for high-Q dielectric cavities[27].

To characterize the success rate of quantum emitter formation the monolayer $WSe_2$ was stamped onto the nanopillar array, but the planar Au layer depicted in **Fig.1a** was not attached, which defines the "uncoupled" case. Optical image analysis reveals that the $WSe_2$ monolayer covers 50 nanopillars excluding sites with missing pillars (**Fig. 2a**). To determine how many nanopillar sites induce quantum emission we have carried out photoluminescence (PL) mapping spectrally selective to the deeply-localized exciton emission (**Fig. 2b**). As anticipated, the bright emission spots follow the lithographically defined locations in 45 out of 50 cases, resulting in a success rate of 90%. The SEM images in **Fig. 2c** verify that the material has been successfully transferred onto the nanopillars without piercing through. As a result of local strain, there are distinct spectral signatures, as shown in **Fig. 2d**. Between nanopillars where the material adheres flat to the substrate an omnipresent and spectrally broad (730-750 nm) defect-related exciton emission band is observed (**Fig.2d** bottom panel). In stark contrast, on nanopillars additional sharp spectral lines appear, which are significantly brighter and energetically shifted to longer wavelength (**Fig.2d** top panel). We focus our study on these deeply-localized emitters created between 750-800 nm (1.55-1.65 eV), as highlighted by the shading in **Fig.2d**. Exemplary spectra for five pillars display between one to seven spectrally isolated emitters (**Fig.2e**), with a mean value of 3 emitters per lithographically defined site over the entire array, that can be easily spectrally filtered. The occurrence histogram in **Fig.2f** shows that strain-induced emitters cluster around 760 nm and taper off towards 800 nm. As a result, quantum emitter formation occurs in a well-defined wavelength range (50 nm) that simplifies emitter-mode coupling, since the spectral width of the plasmonic gap mode (**Fig.1f**) covers the entire occurrence range. We have observed comparable results when transferring monolayer $MoSe_2$ onto the nanocube arrays (see **Supplementary Note 3**), indicating that our approach can be applied to any 2D material.

We consider each observed spectrally sharp peak to be a single quantum emitter, with one example of pronounced single photon antibunching shown in **Fig.3a**. The second-order intensity correlation function $g^2(\tau)$ has a zero-delay time value of $g^2(0)=0.16 \pm 0.03$ taking into account for the timing jitter (**Supplementary Note 4**), as well as a photon recovery time of 3.8 ns, consistent with previous findings for nanobubble-induced[17] as well as pillar-induced quantum emitters[18,19]. In addition, we provide here the first investigation of the quantum emitter coherence time ($T_2$) in $WSe_2$ measured directly in the time domain (**Fig.3b**). The monoexponential decay of fringe visibility yields $T_2 = 3.5 \pm 0.2$ ps. The direct interferometric measurement requires high optical pump power (0.5 mW), which limits $T_2$ due to significant pump-induced dephasing. Alternatively, one can estimate $T_2$ from the spectral linewidth that can be recorded at low pump power (30 nW) where pump-induced dephasing is minimized, resulting in $T_2 = 12$ ps (see **Supplementary Note 5**). These values for 0D excitons are significantly longer than the $T_2$ times reported for delocalized 2D excitons in TMDCs that range from approximately 200 fs-1 ps [28]. Apparently, the demonstrated strain-induced trapping of 2D excitons into 0D confinement potentials within the TMDC material leads to one order of magnitude better protection against pump-induced dephasing.

Emitter-mode coupling in plasmonic nanocavities was often quantified by comparing the coupled emitters with another set of emitters located in reference samples[20,25]. Since the optical properties among individual quantum emitters can vary considerably, statistical approaches are necessary to determine average cavity-coupling properties. Our unique sample design offers the ability to directly compare the optical properties of the same quantum emitter in its coupled and uncoupled state. To this end we devised a measurement scheme depicted in **Fig.4a-b** that first characterizes strain-induced excitons in the presence of the Au nanocubes but without the planar Au mirror, such that the vertical gap mode is not established. This geometry characterizes the "uncoupled" state. After measurements are taken the sample is then inverted and attached in contact with a planar Au mirror to characterize the coupled state. The hyperspectral PL maps of a 3×4 quantum emitter array recorded before (**Fig.4c** top panel) and after (**Fig.4c** bottom panel) introducing the Au mirror show that quantum emitters are significantly brighter when the cavity is formed. All properties discussed in the following are recorded for the most dominant individual quantum emitter located at each site on the array that was spectrally filtered for further investigation. To determine the light enhancement factor EF, the time-integrated intensity of an individual emitter was recorded as a function of excitation power (**Fig.4d**). The intensity increases linearly until saturations sets in, which is the typical hallmark of 0D exciton emission, and in agreement with the observation of single-photon antibunching (**Fig. 4h**). In the saturation region EF = 13, which has contributions from enhanced light extraction, enhanced absorption, and SE rate enhancement. Simulations of the far-field radiation pattern of a dipole emitter determine the light extraction efficiency to be 46% under coupling (**Supplementary Note 1**). Due to the non-resonant excitation scheme with the 532 nm pump laser away from the plasmon resonance one does not expect a significant change in the absorption rate. The observed EF is thus predominantly caused by an enhanced SE rate, i.e. an underlying Purcell effect while enhanced absorption and light collection play a minor role. Taking into account for the limited detection efficiency of the measurement system of 0.21±0.1 % (**Supplementary Note 6**), the single-photon emission rate into the first lens reaches up to 35 MHz.

To directly measure the SE rate enhancement for each individual quantum emitter on the 3x4 array we determined the total rate enhancement factor $\gamma_{on}/\gamma_{off}$ via time-correlated single photon counting (TCSPC), where $\gamma_{on}$ and $\gamma_{off}$ are the rates for the coupled and uncoupled case, respectively. **Figure 4e** shows the TCSPC traces for an exemplary quantum emitter with mono-exponential decay times of $\tau_{off}$ = 5500 ± 45 ps and $\tau_{on}$ = 98 ± 3 ps. The dramatically faster decay time from the coupled emitter yields a rate enhancement of $\gamma_{on}/\gamma_{off}$ = 57. To show the universal coupling we carried out TCSPC experiments for all 12 quantum emitters and found an average lifetime $\tau_{off,av}$ = 4 ± 1.8 ns for the uncoupled case (**Fig 4f**) and $\tau_{on,av}$ = 266 ± 123 ps (**Fig. 4g**) for the coupled case. As anticipated in the design, all 12 sites on the array display a strain-induced quantum emitter featuring a strong Purcell-effect with an average rate enhancement $\gamma_{on}/\gamma_{off}$=15, implying 100% success rate within the selected range.

The measured rate enhancement corresponds only in the case of an emitter with unity QY, i.e. $\eta$=100 % to the Purcell factor $F_P$. For imperfect quantum emitter with low QY that are dominated by NR emission, as it is the case for WSe$_2$ with reports ranging from $\eta$=0.4−4 % [29,30], the underlying $F_P$ is much larger than the measured rate enhancement. For example, at $\eta$=1% and $F_P$=100 the experimentally accessible rate enhancement (intensity enhancement) would just double, since the NR channel dominates. To determine $F_P$ for our case we have directly measured the QY under pulsed excitation in the saturation regime resulting in $\eta_{off}$ = 1.1 ± 0.4% for the uncoupled case and a Purcell-enhanced QY of $\eta_{on}$=14 ± 2% (**Supplementary Note 6**). Note that the latter case also includes the plasmonic metal loss. As is well known, [25] the underlying $F_P$ can in this case be calculated from the product of the measured rate enhancement $\gamma_{on}/\gamma_{off}$ = 57 and the measured QY enhancement $\eta_{on}/\eta_{off}$=13, resulting in $F_P$= 740 for the best case (average $F_P$=272) in **Fig.4d-e** (**Supplementary Note 7**). Such a high $F_P$ implies that the SE coupling factor $\beta=F_P/(1+F_P)$ is unity (99.9%), i.e. the entire SE is coupled to the nanogap mode. Another benefit of coupling to the cavity mode is that the single-photon emission is now purified, as is evident from the antibunching trace in **Fig. 4h** that yields $g^{(2)}(0)$=0.06 ± 0.03 after deconvolution from detector jitter (**Supplementary Note 4**). We note that cavity-enhanced QY is ultimately limited by the metal loss, and can thus be further improved by reducing the gap size to about $\eta$=80% as was experimentally demonstrated for other quantum emitters[23,24,25]. As a result of deterministic coupling, the 12 individual quantum emitters studied of the 3×4 array display an average Purcell enhancement of purified single photons by two orders of magnitude, and a QY enhancement by one order of magnitude.

Finally, we would like to point out that additional rate enhancement of the quantum emitters can be achieved by reducing NR recombination via magnetic brightening. The pronounced asymmetry of the spectral lineshape in **Fig.2e** originates from an underlying doublet fine-structure that is clearly resolvable in high resolution spectra with a zero-field splitting of $\Delta_0$=650 µeV (**Supplementary Note 8**). Corresponding magneto-optical PL spectra reveal a g-factor of g=6.3 ± 0.2 for this emitter. We studied the strain-induced quantum emitters in magnetic fields of up to 9 T, resulting in 35% optical brightening as well as a 36% faster decay rate of $\tau_{on,B}$ = 72 ps at 9 T for the coupled cased (**Supplementary Note 8**). In this case, $F_p$=1050 and $\eta_{on,B}$ = 15 ± 2%. This brightening could potentially be caused by an effective conversion of dark excitons into bright excitons[31]. While a 9 T field is not practical for single photon source device technologies, the magneto-PL finding underpins that further advances of quantum state engineering in 2D materials, for example by forming 2D heterostructures[32] or creating deeper strain-potentials, could lead to further improved QY.

The demonstrated high-yield for Purcell enhancement of single photon generation at lithographically defined locations make the plasmonic nano-gap array approach a promising platform for multi-node quantum optical circuits. The utility of our nanoplasmonic platform reaches beyond the family of TMDC materials and should also be applicable to room-temperature quantum emitters in boron nitride. The small footprint of the demonstrated 3×4 quantum emitter array and directional outcoupling already allows for efficient butt-coupling to optical fibers. When combined with advances in electroluminescence generation from $WSe_2$ [33–35], electrically driven quantum emitter arrays should be within technological reach.

## Methods
**Plasmonic chip fabrication:** The nanocube arrays were fabricated by electron-beam lithography (EBL) using 495 Poly(methyl methacrylate) (PMMA) A4 (MicroChem) that was spin-coated at 2000 rpm onto the substrate. To reduce substrate charging the sapphire wafer was coated with 8 nm aluminum by thermal evaporation before EBL. The side lengths of the individual cubes were defined to be 110 nm, with a height of 90 nm. The samples were subsequently patterned in an Elionix ELS-G100 EBL system and developed in MIBK:IPA for 180 sec. To convert the polymer template into a plasmonic array we deposited a 5 nm Ti adhesion layer and 90 nm Au metal on the samples in an electron beam evaporator (AJA Orion 3-TH) followed by liftoff in Dimethylacetamide solvent at room temperature. Finally, a 2 nm thick layer of $Al_2O_3$ was deposited by atomic layer deposition (ALD). To fabricate the planar Au mirror a 5 nm Ti adhesion layer and a 100 nm Au layer were deposited by slow-rate electron beam evaporation onto an epi-ready sapphire substrate which has a roughness <0.3 nm, and subsequent ALD of 2 nm $Al_2O_3$. This approach avoids detrimental trapping of air-gaps via surface roughness in the planar Au mirror that would otherwise artificially enlarge the plasmonic gap spacing and thus reduce the coupling strength.

**$WSe_2$ exfoliation and transfer:** $WSe_2$ from HQ graphene was mechanically exfoliated onto a viscoelastic stamp (Gel-Film®) using Nitto Denko tape. Thin layers were identified by their optical contrast using an optical microscope and then stamped onto the fabricated arrays where the surface was cleaned and made hydrophilic to improve adhesion between the flake and the substrate. To achieve this, the substrate was submerged in piranha solution for 10 min and rinsed in DI water. Stamping was carried out immediately after surface preparations at an elevated substrate temperature of 60° C to prevent nanobubble formation[17].

**Photoluminescence spectroscopy:** Micro-photoluminescence (µ-PL) measurements were taken inside a closed-cycle cryogen-free cryostat with a 3.8 K base temperature and ultralow vibration (attodry1100 by attocube). Samples were excited with a laser diode operating at 532 nm in continuous wave mode. A laser spot size of about 0.65 µm was achieved using a cryogenic microscope objective with numerical aperture of 0.82. The relative position between sample and laser spot was adjusted with cryogenic piezo-electric *xyz*-stepper while 2D scan images were recorded with a cryogenic 2D-piezo scanner (attocube). Spectral emission from the sample was collected in a multimode fiber, dispersed using a 0.75 m focal length spectrometer, and imaged by a liquid nitrogen cooled silicon CCD camera. Laser stray light was rejected using a 700 nm long-pass edge filter. For time-resolved PL lifetime measurement light from a supercontinuum laser (NKT photonics) operating at 78 MHz repetition rate and 7-ps pulse width was filtered by a 10 nm bandpass centered at 550 nm. TCSPC experiments were carried out with a coincidence counter (SensL) and a fast avalanche photodiode (APD) with a timing jitter of 39 ps (IDQuantique). The system response function was measured sending scattered laser light from the sample surface with the 700 nm filter removed and at an APD count rate that matches the rate of the exciton emission. The second-order correlation function $g^2(\tau)$ was recorded with the same TCSPC setup but in Hanbury-Brown and Twiss configuration where the exciton emission is first sent through a 10 nm bandpass filter followed by a 50:50 multimode fiber splitter before reaching the two APDs.

**Data availability**
The data that support the findings of this study are available from the corresponding author upon request.

**Acknowledgements**
We like to thank Milan Begliarbekov for supporting the EBL process development at the City University of New York Advanced Science Research Center (ASRC) nanofabrication facility. We like to thank Lijun Dai for supporting the design of Figure 1a. S.S. and J.H. acknowledge financial support by the National Science Foundation (NSF) under award DMR-1506711 and DMR-1507423, respectively. S.S. acknowledges financial support for the attodry1100 under NSF award ECCS-MRI-1531237.

**Author contributions**
S.S. and Y.L. designed the experiment. Y.L fabricated the plasmonic chips. G.S. and J.A. performed the layer transfer procedures. Y.L. and G.S. performed the optical experiments and analyzed the data. S.S., J.H., G.S. and Y.L. co-wrote the paper. All authors discussed results and commented on the manuscript.

**Additional information**
The authors declare no competing financial interests. Supplementary information accompanies this paper.

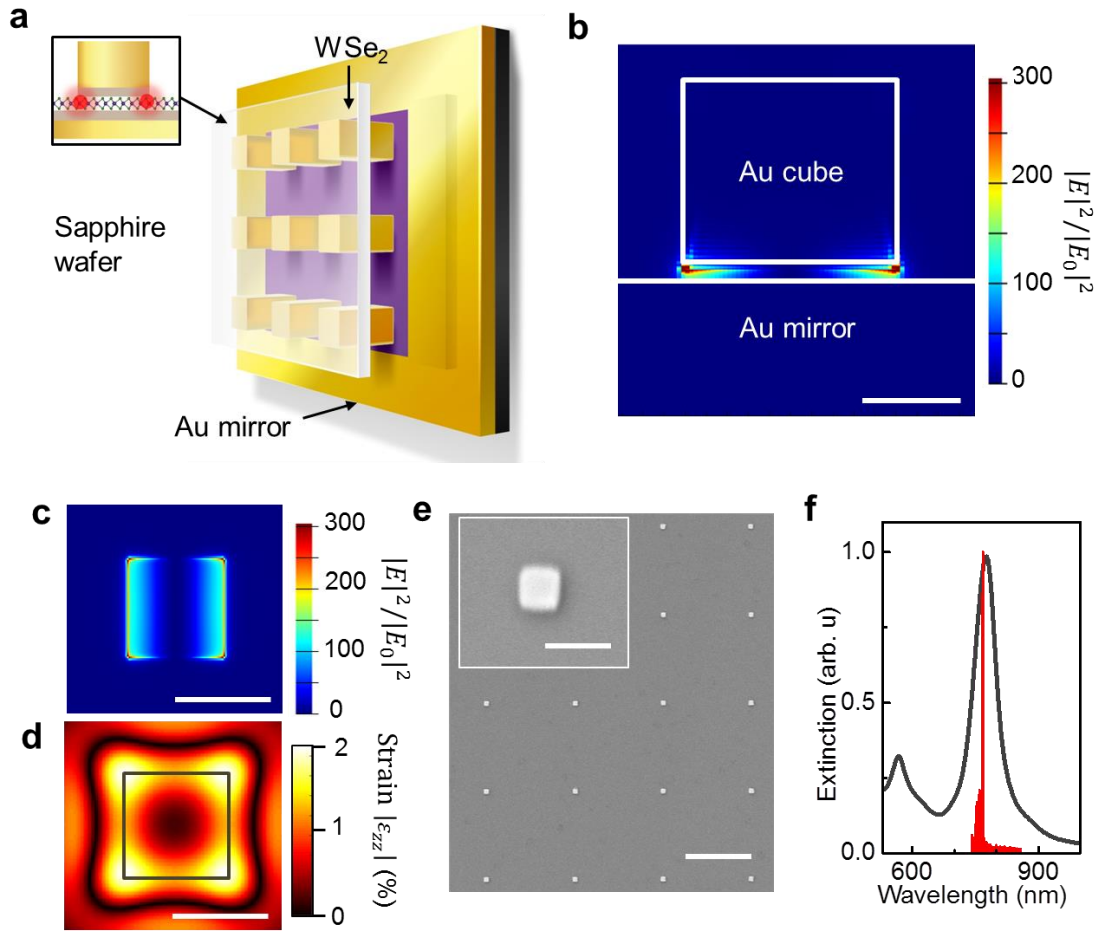

**Fig.1 Overview of sample design enabling deterministic coupling of strain-induced excitons to nanoplasmonic gap modes. a**, Schematic of monolayer (1L) WSe$_2$ coupled to a plasmonic gold nanocube cavity array. The 1L WSe$_2$ is separated from the plasmonic Au cubes and the planar Au layer by a 2 nm Al$_2$O$_3$ spacer layer on each side to prevent optical quenching and short-circuit of the nanoplasmonic gap-mode (depicted with grey shading in the inset). The transparent sapphire wafer allows optical addressing of quantum emitters. **b**, Side view of simulated field enhancement distribution profile illustrating confinement of the plasmon across the vertical gap containing the WSe$_2$. Scale bar: 50 nm. **c**, Top view of simulated field enhancement showing four plasmon hot spots at cube edges. Scale bar: 100 nm. **d**, Simulation of strain profile induced into the WSe$_2$ layer when stamped onto a nanocube indicating highest strain occurs at cube edges coinciding with plasmonic hot spots. Scale bar: 100 nm. **e**, Scanning electron microscope (SEM) image of Au nanocube array made by EBL on sapphire substrate. Scale bar: 2 μm. Inset: zoom into an individual Au nanocube. Scale bar: 200 nm. **f**, Simulated plasmon resonance spectrum (black solid line) together with localized exciton spectrum (red solid line) of a typical quantum emitter directly induced by the Au nanocube into the WSe$_2$ layer illustrating spectral resonance matching.

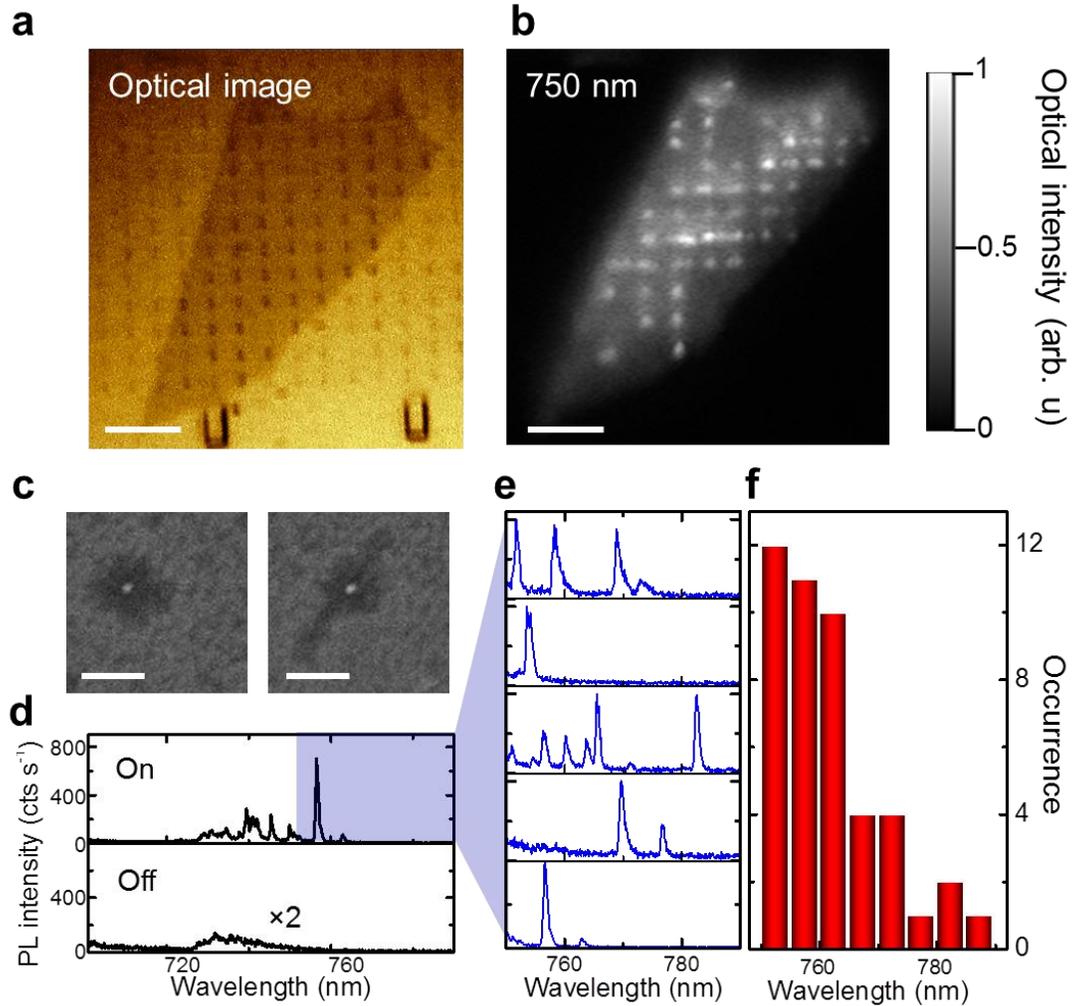

**Fig.2 Optical characterization of strain-induced quantum emitters created by the nanopillar array.**
**a**, Optical image of the WSe$_2$ flake on Au nanopillar array. Scale bar: 5 μm. **b**, Hyperspectral PL map filtered over the spectral range 750-850 nm covering all localized exciton emission. Scale bar: 5 μm. **c**, SEM images of two exemplary locations show that the strained 1L WSe$_2$ material remains elevated for about 100-200 nm around the pillar location (darker electron signal at in-lens detector) before touching the substrate (brighter electron signal). No cracks or damage are visible on the transferred WSe$_2$ layer. Note that round Au cylinders with 100 nm diameter and height are used in this case, i.e. a flat top almost identical to the geometry of the nanocubes. Scale bar: 200 nm. **d**, Comparison of PL spectra when WSe$_2$ is strained over the Au nanopillars (top) and when WSe$_2$ resides on substrate (bottom). **e**, Example normalized PL spectra of isolated quantum emitters at the Au pillar locations. **f**, Histogram of quantum emitter occurrence versus wavelength (i.e. exciton localization depth) recorded from 18 pillar locations.

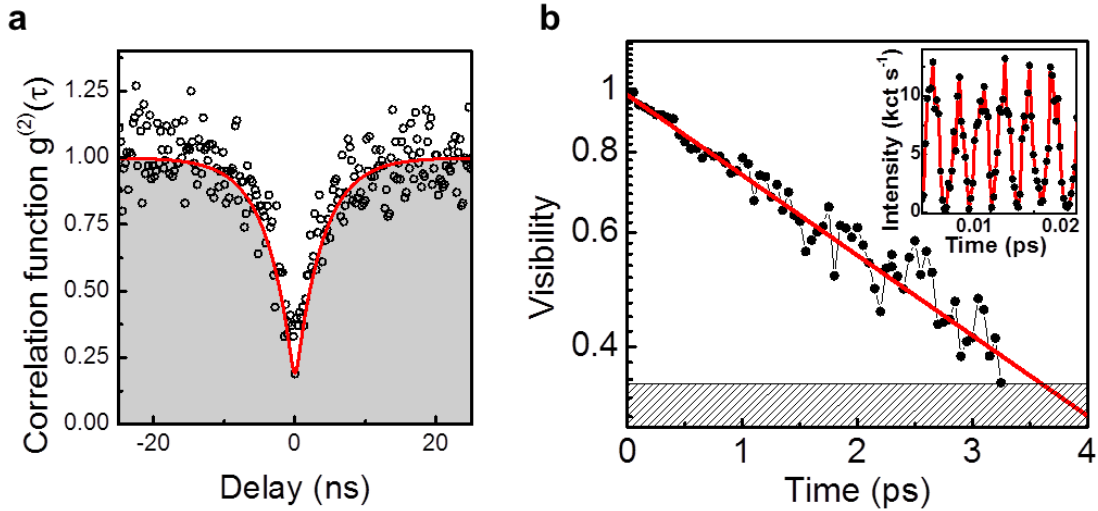

**Fig.3 Quantum light emission and exciton coherence time. a**, Second-order photon correlation function recorded with a Hanbury-Brown Twiss type setup for a strain-induced quantum emitter under non-resonant excitation that displays pronounced single-photon antibunching (black dots). The solid line is a fit to a standard rate-equation analysis for an individual two-level system resulting in a single-photon purity of $g^{(2)}(0) = 0.16 \pm 0.03$ and an exciton recovery time of $\tau = 3.8 \pm 0.2$ ns. **b**, First-order photon correlation function $g^{(1)}(\tau)$ recorded with a Michelson interferometer for a strain-induced quantum emitter. Individual visibility values are calculated by averaging 10 adjacent fringes and are shown as black dots. Red line is fitted to a pure mono-exponential decay function $g^{(1)}(\tau) \sim \exp(-|\tau|/\tau_c))$ revealing a long exciton dephasing time of $T_2 = 3.5 \pm 0.2$ ps. The shaded area illustrates the noise floor. Inset: Zoom into temporally-resolved interference fringes around zero time-delay showing near-unity visibility (V= 0.99).

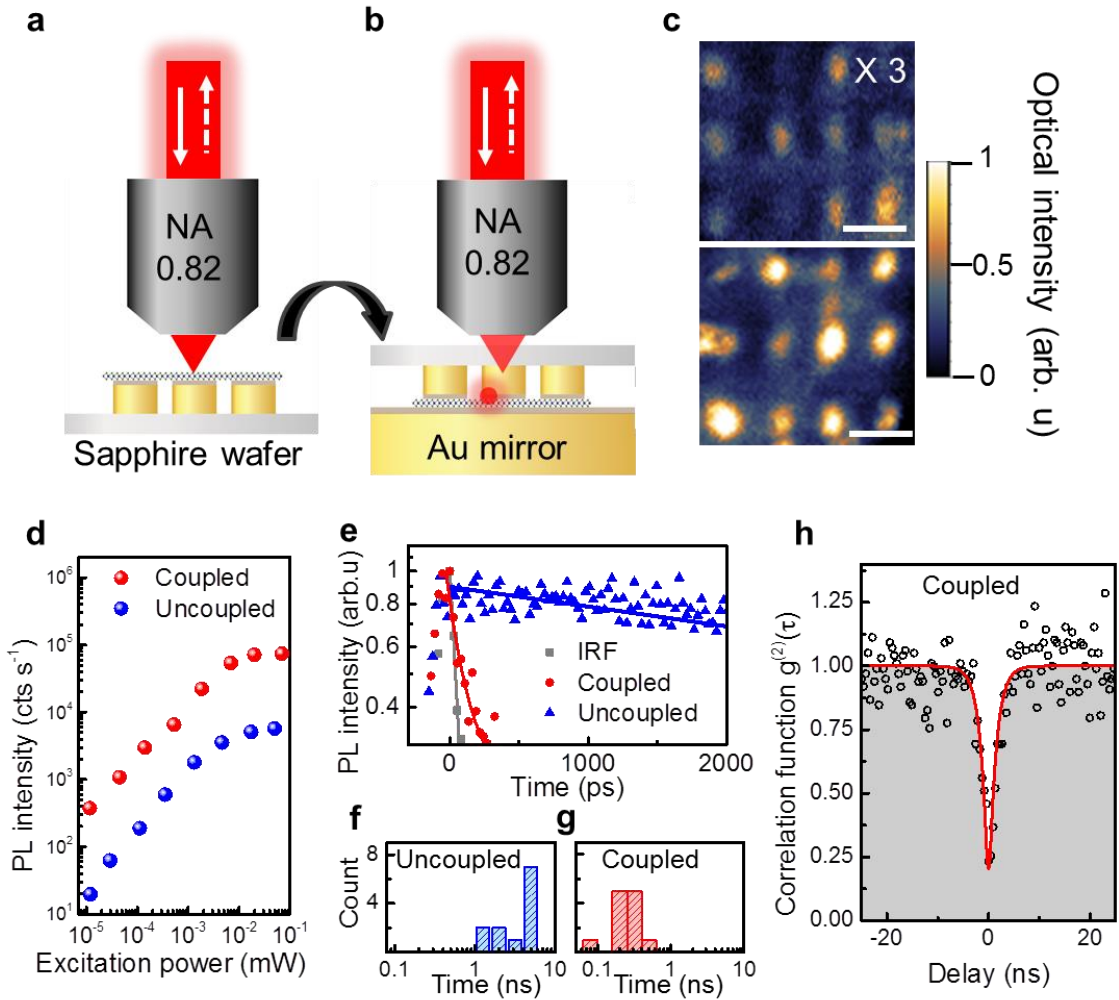

**Fig.4 Quantifying Purcell enhancement of plasmonically coupled quantum emitters**. **a-b,** Schematic of the sample configuration to measure PL intensity and spontaneous emission (SE) rate from each strain-induced quantum emitter before (**a**) and after (**b**) formation of the plasmonic nanocavity by introducing the planar Au mirror. **c**, Hyperspectral PL map of a 3×4 quantum emitter array recorded through the sapphire side and taken before (top panel) and after (bottom panel) the cavity was closed with the planar Au mirror. Note the scan in the top panel is magnified three-fold to illustrate that all 12 sites contain strain-induced quantum emitter emission. Scale bar: 2 μm. **d**, Integrated PL intensity of an exemplary quantum emitter as a function of excitation power. Blue circles recorded under configuration (**a**) and red circles for the same emitter location under configuration (**b**). **e**, SE lifetime measurements recorded by TCSPC at 40 μW excitation power. Gray squares: System response for back-reflected laser light. Solid gray line: Mono-exponential fit representing the system response (65 ps). Blue triangles: isolated quantum emitter without the cavity. Solid blue line is a fit with a mono-exponential decay time of $\tau_{off}$ = 5500 ± 40 ps. Red circles: same emitter recorded when coupled to the nanocavity. Red solid line: fit with $\tau_{on}$ = 98 ± 3 ps. PL intensity is shown on logarithmic scale. **f**, **g** Histogram of lifetimes for the 12 isolated quantum emitter from the 3×4 array measured without the cavity mirror (**f**) and when coupled (**g**). **h**, Purified single-photon emission through cavity-coupling resulting in $g^{(2)}(0)$ = 0.06 ± 0.03 after deconvolution from detector jitter.

# Supplementary Note 1. Numerical simulations

To guide the structural design of the plasmonic Au nanocubes before nanofabrication we have carried out finite-difference time-domain (FDTD) simulations using Lumerical FDTD Solutions software package to determine the physical dimensions that result in plasmon resonance energies that match with the known emission energy of the exciton emission. In addition, the FDTD simulations can determine expected Purcell enhancement, light collection efficiency, and local heat maps. A total-field scatter-field source was used to excite the plasmonic structures and a simulation region with 1 nm uniform mesh size in $x$-, $y$-, and $z$- directions was defined. The dielectric constant data for Au measured by Johnson and Christy[1] was used to model the plasmon resonances. The dielectric function for the $Al_2O_3$ was modeled from the measurement of Palik[2]. Frequency-domain field and power monitors were placed near the sample surface to record the near-field electrical field intensity. To determine the plasmon resonance wavelength we varied the side size of the Au nanocube leaving the height fixed to 90 nm. The vertical gap between the nanocube and planar Au mirror is set to 5nm. Given that the center of the plasmon mode should be around 780 nm to support optical emission from isolated quantum emitters we estimated that the optimized side length of the nanocube structure should be 110 nm.

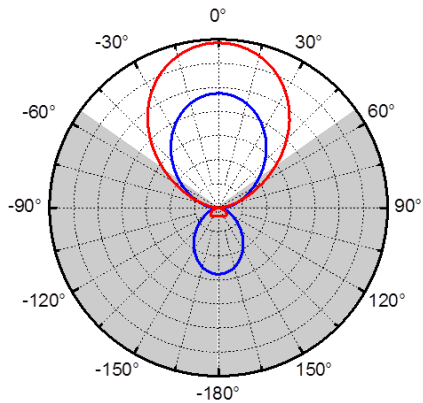

**Supplementary Figure 1. Light collection efficiency calculation**. FDTD calculated far-field projection pattern for an uncoupled (blue line) and coupled (red line) quantum emitter.

In addition, the light collection efficiency of coupled and uncoupled dipole emitters was determined by FDTD simulations resulting in the far-field projection pattern shown in **Supplementary Fig. 1**. For coupled quantum emitter simulation, a dipole source has been placed on the structure shown in **Fig.1a**. The far-field pattern is generated using a frequency-domain field power monitor. For the uncoupled case the Au mirror was not present. The objective lens has a numerical aperture NA = 0.82 resulting in a light collection angle of ±55°. All results are calibrated with the dipole source power to extract the collection efficiency. The calculated collection efficiency for the plasmonic cavity was found to be $\eta_c$ = 83%, while without the cavity the Au nanocube alone yields $\eta_0$ = 57%. The additional enhanced light extraction compared to the uncoupled emitter on the Au nanocube is thus $\varepsilon$ = 1.46.

# Supplementary Note 2. Strain map calculation

To calculate the strain profile of the $WSe_2$ flake located on the nanocube we used infinitesimal strain theory, i.e. a classical continuum model that is commonly used for structural stress analysis[3]. The height profile $H(x, y)$ of the flake was modelled by assuming the flake rests flat on the top of the nanocube and the bending occurs at the cube edge and at the substrate where the flake is supported:

$$H(x,y) = \frac{1}{16} H_0 \left[ \text{erf}\left(\frac{x+\frac{l}{2}}{s_x}\right) + 1 \right] \left[ \text{erf}\left(\frac{y+\frac{l}{2}}{s_y}\right) + 1 \right] \quad (1)$$

where $H_0$ is the height of the nanocube, $l$ is the side length of the nanocube, and $s_x$, $s_y$ are the depression parameters along x and y directions. By modeling the WSe$_2$ flake as a thin elastic sheet, one can calculate the strain by applying $H(x,y)$ into $|\varepsilon_{zz}| = \left| \frac{vd}{1-v} \left( \frac{\partial^2 H(x,y)}{\partial x^2} + \frac{\partial^2 H(x,y)}{\partial y^2} \right) \right|$, where d = 0.7 nm is the monolayer thickness and $v = 0.2$ is the Poisson ratio. The strain map calculated in this way is shown in **Fig.1d**.

### Supplementary Note 3. Strain-induced quantum emitters in MoSe$_2$ monolayers

Apart from WSe$_2$, strain-induced quantum emitters have also been created in WS$_2$ [4], and recently also in hBN [5]. Here we add evidence that they can also be created in MoSe$_2$. As shown in **Supplementary Fig. 2a**, imaging the sample surface with laser stray light reveals a transferred monolayer of MoSe$_2$ onto a plasmonic Au nanocube array with the same aspect ratio as the data for WSe2 presented in the main text. When recorded as a hyperspectral PL image filtered around 800 nm one can identify the nanocube lattice through the enhanced defect-related exciton emission similar to the findings on monolayer WSe$_2$, indicating that PL emission is strain-induced (**Supplementary Fig. 2b**). However, since MoSe$_2$ is a direct band-gap semiconductor at low temperatures, the 2D exciton emission still dominates the low-temperature emission spectra, giving rise to an omnipresent background on which the sharp quantum emitters ride, as depicted in three exemplary spectra in the **Supplementary Fig. 2c**. This background emission is detrimental to the observation of single photon antibunching and thus limits the usefulness of MoSe$_2$ as a quantum emitter material, unless ways are found to create significantly deeper localization potential, e.g. by optimizing the aspect ratio for the stressor array. The advantage of WSe$_2$ is that it undergoes a transition to an indirect-gap semiconductor at cryogenic temperatures resulting in significantly less background and thus higher single-photon purity.

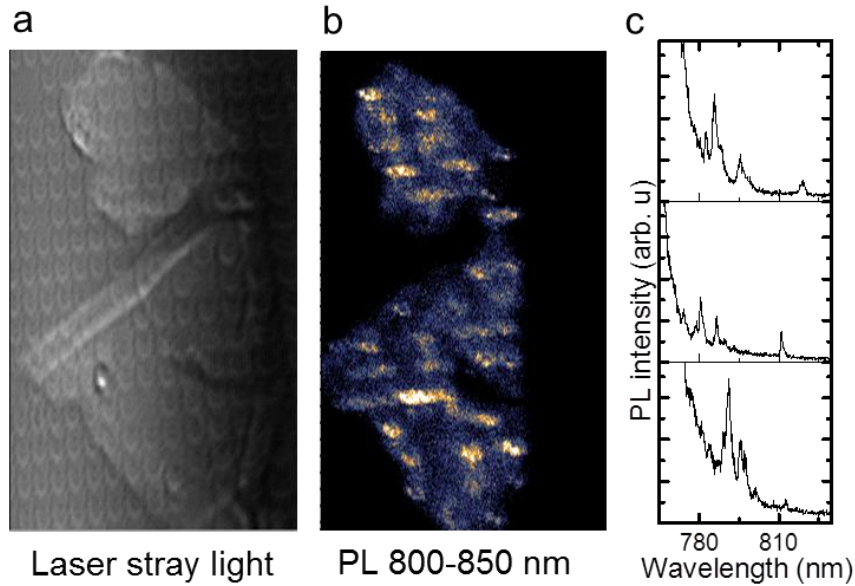

**Supplementary Figure 2. Strain-induced quantum emitters in monolayer MoSe$_2$. a**, Laser stray-light image of transferred MoSe$_2$ flake on Au nanocube array. **b**, Corresponding hyperspectral mapping of PL emission filtered in the 800–850 nm wavelength range where spectrally sharp lines occur. **c**, Exemplary spectra recorded at three different Au nanocube locations showing quantum emitters overlap with a broad background. Data are recorded at 3.8 K.

# Supplementary Note 4. Deconvolution of system jitter from second-order intensity correlation functions

The degree of antibunching in the measured $g^{(2)}(\tau)$ histograms from the Hanbury-Brown Twiss setup are affected by the timing jitter $\sigma$ of the detection system. To determine the jitter of the Perkin-Elmer APDs we have sent laser stray light (pulse width 5 ps) to the APDs, resulting in a TCSPC decay trace that fits mono-exponential to a decay time of 220 ps. The combined system jitter of two detectors is given by $\sqrt{2}\cdot$ 220 ps = 311 ps. Adding the timing jitter of the coincidence electronics of 27 ps by error propagation results in $\sigma$=312 ps for the system response. The system jitter plays a minor role for the uncoupled emitters that have an antibunching recovery time that is with 3.8 ns more than an order of magnitude larger than $\sigma$. It does however significantly affect the $g^{(2)}(\tau=0)$ values of the Purcell-enhanced cases. To illustrate this effect, we plot in **Supplementary Fig. 4a** the response of an ideal 2-level system at a fast recovery time of 200 ps for various values of $\sigma$. While the absence of jitter ($\sigma$=0 ps) produces perfect antibunching with $g^{(2)}(\tau=0) = 0$, our timing jitter of 312 ps diminishes the antibunching to $g^{(2)}(\tau=0) = 0.7$, i.e. almost unresolved. In order to correct for the system jitter we fit the normalized histogram of the coincidence counts with the correlation function $g^{(2)}(\tau) = 1 - a\, e^{-|\tau|/\tau_{rec}}$ convolved with Gaussian distribution error function $\frac{1}{\sqrt{2\pi\sigma^2}}e^{-\tau^2/(2\sigma^2)}$, where $\tau_{rec}$ is the varying exciton recovery time and $\sigma$= 312 ps. Two convolution examples for relatively fast recovery times of 470 ps and 250 ps are shown in **Supplementary Fig. 4b-c**, resulting in deconvolved values of $g^{(2)}(0) = 0.05 \pm 0.03$ and $g^{(2)}(0) = 0.08 \pm 0.04$, respectively. Compared to the uncoupled cases with typical values of $g^{(2)}(0) = 0.16$ the coupled emitters display significantly cleaner antibunching due to the mode filtering effect provided by the gap-mode that suppresses noise photons[6–8].

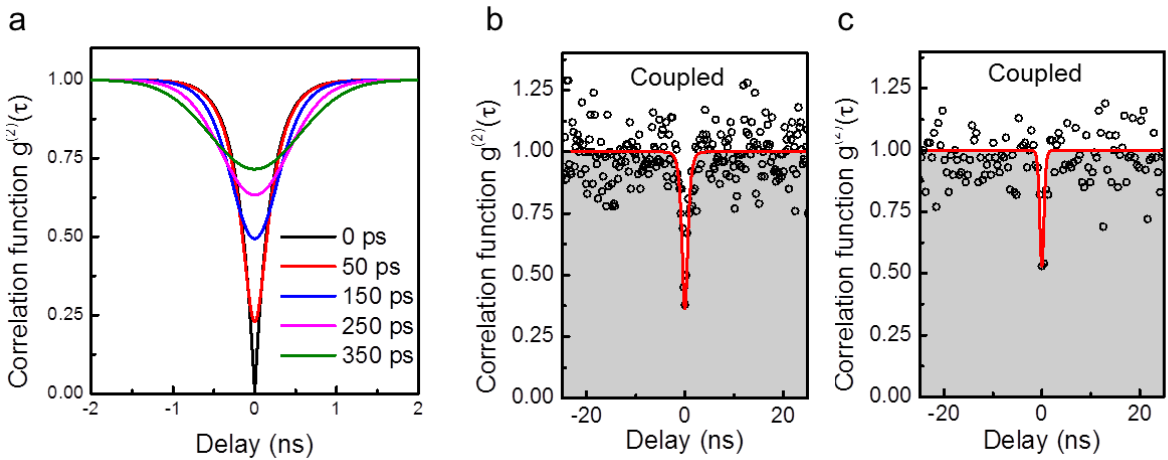

**Supplementary Figure 3. Analysis of second-order correlation traces in the presence of detector jitter. a,** Black solid line: Correlation function for an ideal two-level system with $g^{(2)}(0)$=0 and a photon recovery time of 200 ps in the absence of detector jitter. Colored solid lines: Convoluted response functions for detector jitter varied from 50 ps (red line) to 350 ps (green line). **b,** Convoluted function of emitter fitting to a recovery time of of 470 ps resulting in $g^{(2)}(0) = 0.05 \pm 0.03$. **c,** Convoluted function of emitter with recovery time 250 ps resulting in $g^{(2)}(0) = 0.08 \pm 0.04$.

# Supplementary Note 5. Spectral linewidth deconvolution procedure

Given that the exciton emission from the strain-induced quantum emitters has a spectral linewidth (FWHM) near the resolution limit of the detection system one needs to deconvolve the spectrometer response function. To this end, we have first measured the system response of the spectrometer (1200 groove/mm grating, focal length 0.75 m, slit width setting 20 µm) by sending laser light that is back-reflected from the sample surface through the same optical path as the exciton emission is sent. A single

Gaussian fit to the system response yields FWHM = 68 μeV at 780 nm (**Supplementary Fig. 4a**), with $R^2=0.995$. Using mathematical deconvolution with a Voigt function one can largely eliminate the system response and can determine the underlying FWHM of the quantum emitter that is assumed to be a single Lorentzian, particularly at low pump power where pump-induced dephasing is minimized. The Voigt deconvolution procedure is given by:

$$y = y_0 + A \frac{2\ln2}{\pi^{3/2}} \frac{\omega_0}{\omega_{\text{IRS}}} \int_{-\infty}^{\infty} \frac{e^{-t^2}}{(\sqrt{\ln2}\frac{\omega_0}{\omega_{\text{IRS}}})^2 + (2\sqrt{\ln2}\frac{x-x_c}{\omega_0} - t)^2} dt, \quad (2)$$

where $y_0$, $\omega_0$, $\omega_{\text{IRS}}$, $x_c$ are background counts, FWHM of the emitter spectrum, FWHM of the instrument response, and the resonance frequency, respectively. **Supplementary Fig. 4b** shows an example of the deconvolution analysis for lowest pump powers in the experiment that still maintains a large signal to noise ratio (SNR=32) such that the deconvolution procedure yields statistically significant values. As a result, we find a FWHM for the exciton emission of 121 ± 4 μeV for the dominant component in the Zeeman doublet.

Using the well-known relation $\text{FWHM} = 2\Gamma = \frac{2\hbar}{T_2} = \frac{\hbar}{T_1} + \frac{2\hbar}{T_2^*}$, one can estimate the $T_2$ coherence time of the exciton emission to be at least 12 ± 1 ps as the lower bound.

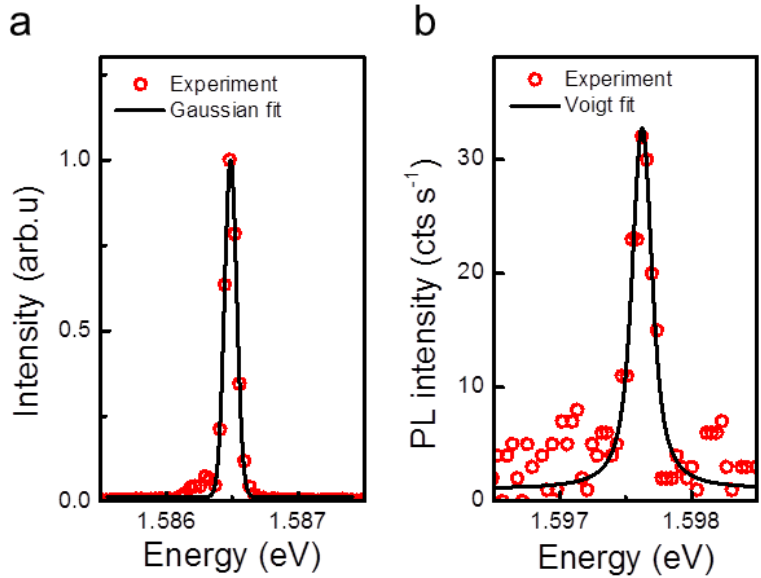

**Supplementary Figure 4. Deconvolution procedure to determine the FWHM of the exciton emission. a**, System response function for 780 nm laser light that fits with a single Gaussian. **b**, The localized exciton emission (red circles) was taken at an excitation pump power of 30 nW. The black solid lines are a fit to the Voigt function defined in Supplementary Equation 2 that includes the Gaussian system response shown in (a). Spectrum was recorded at 3.8 K.

## Supplementary Note 6. External collection efficiency and quantum yield

For a single photon source driven under pulsed laser excitation, the underlying quantum yield of the quantum emitter can be simply calculated from the measured number of photons emitted in the saturation-regime normalized to the laser repetition rate[9,10]. In the ideal case of unity quantum yield, each laser pulse generates one photon and the measured count rate must match the laser repetition rate, if corrected for the detection loss in the experimental light detection system. We have calibrated the losses following our previous work on quantum yield determination from individual carbon nanotubes[11], but here for 780 nm laser light sent to each component. **Supplementary Table 1** lists the external collection efficiency when light is detected with an APD as well as when light is recorded with a CCD camera and spectrometer, resulting in 1% and 0.2% respectively.

|  | APD | CCD/spectrometer |
|---|---|---|
| VISIR cryo objective transmission | 80% | 80% |
| Cryostat optics | 50% | 50% |
| MM fiber coupling | 80% | 80% |
| 10 nm bandpass filter | 26% | 26% |
| Long pass filter | 85% | 85% |
| APD quantum efficiency | 47% |  |
| MM fiber in-coupling |  | 90% |
| 300 gr/mm grating |  | 37% |
| Al-coated spectrometer mirrors $\times$ 3 |  | $(71\%)^3 = 35\%$ |
| CCD camera quantum efficiency |  | 80% |
| Sapphire substrate with 2 nm Al | 32% | 32% |
| **Total collection efficiency** | $1 \pm 0.3$ % | $0.21 \pm 0.1$ % |

**Supplementary Table 1**. **Transmission/reflection losses or yield of optical components.** All values are extracted from experimental measurements using 780 nm laser light except the avalanche photodiode and CCD detector efficiencies which are taken from the manufacturer datasheet. Note that missing entries mean that the specific component is not part of the optical pathway.

To confirm that single-photon emission occurs for strain-induced quantum emitters under pulsed excitation we used an NKT supercontinuum source with a repetition rate of 78 MHz (13 ns) and pulse width of 5 ps that was spectrally filtered through a 10 nm bandpass centered at 590 nm. The resulting second-order photon correlation trace in **Supplementary Fig. 5a** shows pronounced antibunching that is still present when driven near saturation. Note that there is an additional admixing of a continuous component between clock states superposed on top of the pulsed response that also displays pronounced photon antibunching, clearly identifying it as quantum emitter related. It originates from the underlying ns long lifetime component of the exciton emission in the TCSPC experiment that remains after the fast decay component has depleted the emission to about 20-30% of its initial value (see **Fig.4e**). **Supplementary Figure 5b** shows the corresponding emitter count rate as a function of pump power. In the saturation regime the uncoupled emitter displays a 30 kHz single photon count rate on the APD that corrsponds to a collection rate of 960 kHz when corrected for the detetction efficiency. Calibrated to the 78 MHz laser repetition rate the quantum yield of the uncoupled quantum emitter is $\eta_{off} = 1.1 \pm 0.4\%$, which is in the range of previous findings utilizing references dyes for calibration[12,13]. Similarly, the coupled quantum emitter rate of 10.6 MHz into the first lens results in a Purcell-enhanced quantum yield of $\eta_{on} = 14 \pm 2\%$.

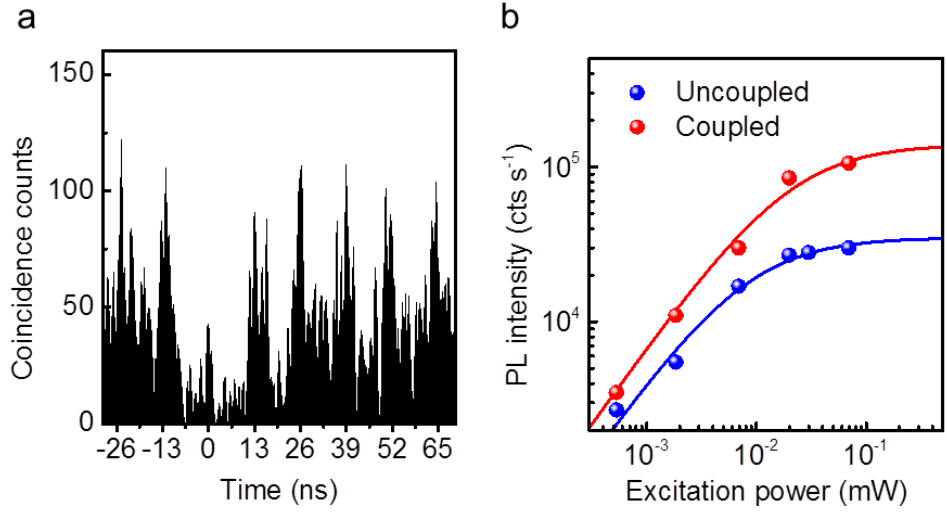

**Supplementary Figure 5. Photon statistics and power dependent PL intensity under pulsed excitation. a,** Second-order correlation function $g^{(2)}(\tau)$ recorded near the saturation regime at $P_{exc}$ = 80 µW excitation power under pulsed excitation demonstrating pronounced single photon antibunching with $g^{(2)}(0)$ = 0.37 ± 0.08. **b,** Integrated PL intensity from uncoupled (blue dots) and coupled (red dots) quantum emitters. Blue solid line and red solid line are theoretical fits following a two-level rate equation model, respectively.

## Supplementary Note 7. Derivation of Purcell factor and quantum yield of WSe₂ quantum emitters coupled to plasmonic nanocavities in the presence of metal loss

The total decay rate through each decay channel for strain-induced quantum emitters in monolayer WSe₂ without coupling to plasmonic nanocavities is $\gamma_{off} = \gamma_R + \gamma_{NR}$, where $\gamma_R$ and $\gamma_{NR}$ are the radiative and nonradiative decay rates. The metal loss rate $\gamma_M$ is an additional decay channel introduced by coupling to plasmonic nanocavities. Therefore, the total decay rate $\gamma_{on}$ of the coupled emitter can be expressed as $\gamma_{on} = (F_p + 1)\gamma_R + \gamma_{NR} + \gamma_M$, where $F_p$ is the Purcell factor that enhances the radiative rate[11]. The ratio of these two rates given by

$$\frac{\gamma_{on}}{\gamma_{off}} = \frac{(F_p+1)\gamma_R + \gamma_{NR} + \gamma_M}{\gamma_R + \gamma_{NR}} \qquad (3)$$

is the quantity that is directly determined in the time-resolved measurements presented in **Fig. 4e**. The experimentally measured rate enhancement of $\gamma_{on}/\gamma_{off}$ = 57 is remarkable given that the starting point is a low quantum yield of only 1%, implying that a significant Purcell enhancement is required to catch up with $\gamma_{NR}$. Even at an order of magnitude enhancement of $\gamma_R$ due to the Purcell effect, the total measured rate $\gamma_{on}$ increases only by a few percent, while $F_p$=100 would just double the measured rate $\gamma_{on}$. In order to determine the underlying Purcell factor from the measured rates we note that quantum yield $\eta$ is defined by the ratio of the number of photons emitted to the number of photons absorbed:

$$\eta_{off} = \frac{\gamma_R}{\gamma_R + \gamma_{NR}} \qquad (4)$$

$$\eta_{on} = \frac{(F_p+1)\gamma_R}{(F_p+1)\gamma_R + \gamma_{NR} + \gamma_M} \qquad (5)$$

The cavity-enhanced quantum yield enhancement is thus given by:

$$\frac{\eta_{\text{on}}}{\eta_{\text{off}}} = \frac{(F_p+1)(\gamma_R+\gamma_{NR})}{(F_p+1)\gamma_R+\gamma_{NR}+\gamma_M} \quad (6)$$

Combining (3) and (6) leads to the expression for the metal-loss corrected Purcell factor $F_p$ given by:

$$F_p = \frac{\gamma_{\text{on}}}{\gamma_{\text{off}}}\frac{\eta_{\text{on}}}{\eta_{\text{off}}} - 1 \quad (7)$$

One can thus simply use the product of the cavity-enhanced quantum yield from the experiments described in **Supplementary Note 6**, and the experimentally measured rate enhancement from the time-resolved TCSPC experiment in **Fig.4** to determine the underlying $F_P$ values, as reported in the main text.

Note that the Purcell enhanced emitter has also a cavity-enhanced quantum yield $\eta_{\text{on}}$ that can be significantly larger compared to the uncoupled emitters, since the Purcell enhanced radiative rate catches up with all nonradiative losses in the system. However, ultimately $\eta_{\text{on}}$ cannot reach unity in plasmonic systems since the metal loss cannot be entirely avoided. At best, for high enough Purcell factors, the cavity-enhanced quantum yield saturates at values determined by the metal loss rate $\gamma_M$. The measured $\eta_{\text{on}} = 14 \pm 2\%$ is quite close to the theoretical expectations from FDTD simulations that result in $\eta_{\text{on}} = 16\%$ from the branching ratio between emitter emission rate and $\gamma_M$, which is solely a property of the plasmonic nanocavity.

| Parameter | Definition |
| --- | --- |
| $\gamma_{\text{off}}$ | total decay rate of quantum emitter without coupling to the cavity |
| $\gamma_{\text{on}}$ | total decay rate of quantum emitter coupled to plasmonic cavity |
| $\gamma_R$ | radiative decay rate of quantum emitter without coupling to the cavity |
| $\gamma_{NR}$ | non-radiative decay rate of quantum emitter without coupling to the cavity |
| $\gamma_M$ | metal loss rate |
| $F_p$ | Purcell factor |
| $\eta_{\text{off}}$ | quantum yield of quantum emitter without coupling to the cavity |
| $\eta_{\text{on}}$ | quantum yield of quantum emitter coupled to plasmonic cavity |

**Supplementary Table 2. Definitions of all parameters involved in the derivation of Purcell factor and quantum yield of strain-induced quantum emitters in monolayer WSe$_2$ coupled to plasmonic nanocavities.**

## Supplementary Note 8. Magneto-PL properties of strain-induced quantum emitters

To investigate the magneto-PL properties of strain-induced quantum emitters in monolayer WSe$_2$ we applied the magnetic field parallel to the k-vector of the incident laser (Faraday configuration). The spectra in **Supplementary Fig. 6a** clearly resolve the doublet structure underlying most strain-induced quantum emitters. This particular case displays a zero-field splitting Δ$_0$ of 650 μeV, which was suggested to result from the electron-hole spin-exchange interaction in an underlying anisotropic strain field[14]. With increasing magnetic field up to 9 T the two components split further apart, while their oscillator strength exchanges to favor the low energy branch at high fields (**Supplementary Fig. 6b**). The Zeeman splitting was analyzed to determine the g-factor using the well know relation: $\Delta_B = \sqrt{(\Delta_0)^2 + (\mu_B g B)^2}$ , where $\mu_B$ is the Bohr magneton and g is the exciton g-factor, revealing g = 6.3 ± 0.2 (**Supplementary Fig. 6c**). **Supplementary Fig. 6d** shows the intensity versus pump power for a coupled emitter comparing the case of zero field to 9 T. At highest pump powers the emitter intensity integrated over both Zeeman components is 35% higher at 9 T as compared to the zero-field case, indicative of magnetic brightening of the quantum emitter. Likewise, **Supplementary Fig. 6e** shows that intensity brightening is accompanied by a 36 % faster decay rate of $\tau_{on,B}$ = 72 ps as compared to the zero field case.

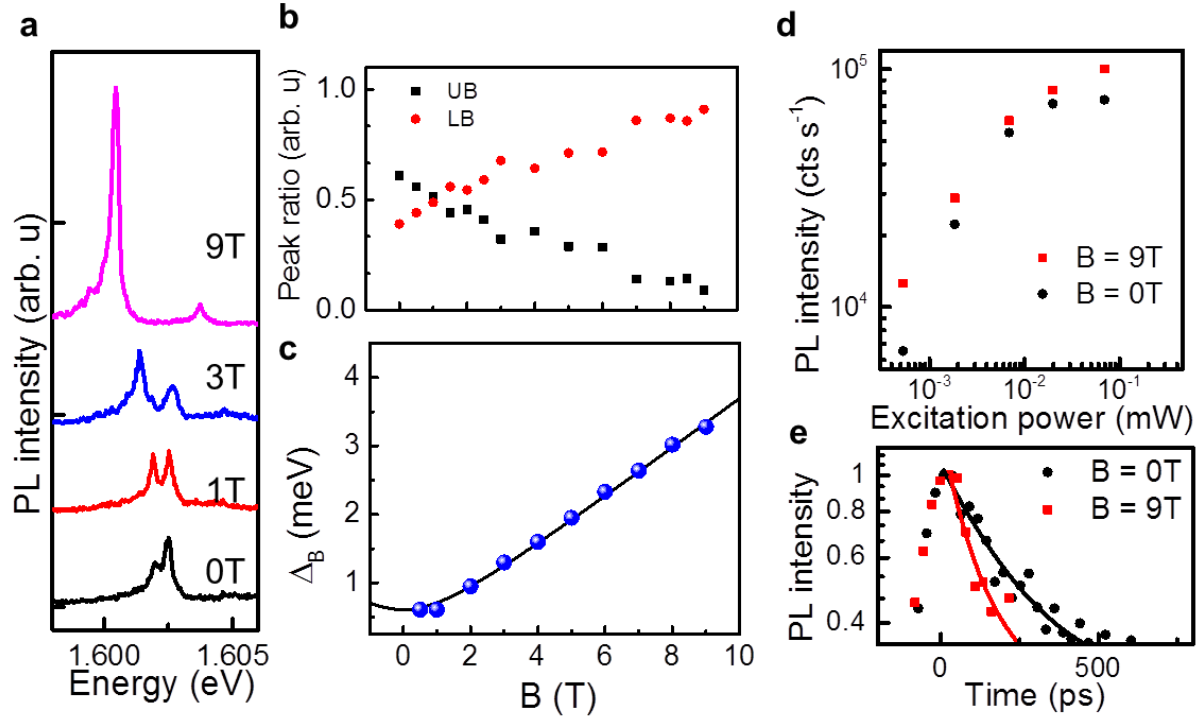

**Supplementary Figure 6. Magnetic brightening**. **a,** Exemplary photoluminescence spectra recorded in Faraday configuration for varying magnetic field strength. **b,** Magnetic field dependence of peak intensity ratio of the upper branch (UB: black squares) and lower branch (LB: red dots). **c,** Zeeman energy splitting Δ$_B$ as a function of magnetic field. **d,** Integrated PL intensity of the same emitter presented in **Fig. 4** as a function of excitation power recorded at zero magnetic field (black dots) and B = 9 T (red squares). **e,** Corresponding TCSPC measurement resulting in a 36% faster decay rate at B = 9 T. Data are recorded at 3.8 K.